\documentclass[10pt,letterpaper]{article}
\usepackage[T1]{fontenc}
\usepackage{makecell}
\usepackage{enumitem}
\usepackage{lscape}
\usepackage{pifont}
\newcommand{\xmark}{\ding{55}}
\newcommand{\cmark}{\ding{51}}
\usepackage[top=0.85in,left=2.75in,footskip=0.75in]{geometry}

\usepackage{amsmath,amssymb}

\usepackage{changepage}

\usepackage[utf8x]{inputenc}

\usepackage{textcomp,marvosym}

\usepackage{cite}

\usepackage{nameref,hyperref}

\usepackage[table]{xcolor}

\usepackage{array}

\newcolumntype{+}{!{\vrule width 2pt}}

\newlength\savedwidth

\raggedright
\usepackage[aboveskip=1pt,labelfont=bf,labelsep=period,justification=raggedright,singlelinecheck=off]{caption}

\makeatletter
\renewcommand{\@biblabel}[1]{\quad#1.}
\makeatother

\usepackage{lastpage,fancyhdr,graphicx}
\usepackage{epstopdf}
\fancyheadoffset[L]{2.25in}
\fancyfootoffset[L]{2.25in}
\begin{document}
\vspace*{0.2in}

\begin{flushleft}
{\Large
\textbf\newline{Robust Framework for COVID-19 Identification from a Multicenter Dataset of Chest CT Scans} 
}
\newline
\\
Sadaf Khademi\textsuperscript{1\Yinyang},
Shahin Heidarian\textsuperscript{2\Yinyang},
Parnian Afshar\textsuperscript{1},
Nastaran Enshaei\textsuperscript{1},
Farnoosh Naderkhani\textsuperscript{1},
Moezedin Javad Rafiee, MD\textsuperscript{3},
Anastasia Oikonomou, MD\textsuperscript{4},
Akbar Shafiee, MD\textsuperscript{5},
Faranak Babaki Fard, MD\textsuperscript{6},
Konstantinos N. plataniotis\textsuperscript{7},
Arash Mohammadi\textsuperscript{1*},
\\
\bigskip
\textbf{1} Concordia Institute for Information Systems Engineering, Concordia University, Montreal, Canada
\\
\textbf{2} Department of Electrical and Computer Engineering, Concordia University, Montreal, QC, Canada
\\
\textbf{3} Department of Medicine and Diagnostic Radiology, McGill University, Montreal, QC, Canada
\\
\textbf{4} Department of Medical Imaging, Sunnybrook Health Sciences center, Toronto, Canada
\\
\textbf{5} Department of Cardiovascular Research, Tehran Heart Center, Cardiovascular Diseases Research Institute, Tehran University of Medical Sciences, Tehran, Iran
\\
\textbf{6} Faculty of Medicine, University of Montreal, Montreal, QC, Canada
\\
\textbf{7}   Department of Electrical and Computer Engineering, University of Toronto, Toronto, Canada
\\
\bigskip

%
%
\Yinyang These authors contributed equally to this work.

%
%
%

* arash.mohammadi@concordia.ca

\end{flushleft}
\section*{Abstract}
The main objective of this study is to develop a robust deep learning-based framework to distinguish COVID-19, Community-Acquired Pneumonia (CAP), and Normal cases based on volumetric chest CT scans, which are acquired in different imaging centers using different scanners and technical settings. We demonstrated that while our proposed model is trained on a relatively small dataset acquired from only one imaging center using a specific scanning protocol, it performs well on heterogeneous test sets obtained by multiple scanners using different technical parameters.  We also showed that the model can be updated via an unsupervised approach to cope with the data shift between the train and test sets and enhance the robustness of the model upon receiving a new external dataset from a different center. More specifically, we extracted the subset of the test images for which the model generated a confident prediction and used the extracted subset along with the training set to retrain and update the benchmark model (the model trained on the initial train set). Finally, we adopted an ensemble architecture to aggregate the predictions from multiple versions of the model. For initial training and development purposes, an in-house dataset of 171 COVID-19, 60 CAP, and 76 Normal cases was used, which contained volumetric CT scans acquired from one imaging center using a single scanning protocol and standard radiation dose. To evaluate the model, we collected four different test sets retrospectively to investigate the effects of the shifts in the data characteristics on the model’s performance. Among the test cases, there were CT scans with similar characteristics as the train set as well as noisy low-dose and ultra-low dose CT scans. In addition, some test CT scans were obtained from patients with a history of cardiovascular diseases or surgeries.  This dataset is referred to as the ``SPGC-COVID'' dataset. The entire test dataset used in this study contains 51 COVID-19, 28 CAP, and 51 Normal cases. Experimental results indicate that our proposed framework performs well on all test sets achieving  total accuracy of 96.15\% (95\%CI : [91.25 - 98.74]), COVID-19 sensitivity of 96.08\% (95\%CI : [86.54 - 99.5]), CAP sensitivity of 92.86\% (95\%CI : [76.50 - 99.19]), Normal sensitivity of 98.04\% (95\%CI : [89.55 - 99.95]) while the confidence intervals are obtained using the significance level of 0.05. The obtained AUC values (One class vs Others) are 0.993 (95\%CI : [0.977 - 1]), 0.989 (95\%CI : [0.962 - 1]), and 0.990 (95\%CI : [0.971 - 1]) for COVID-19, CAP, and Normal classes, respectively. The experimental results also demonstrate the capability of the proposed unsupervised enhancement approach to improve the performance and robustness of the model when being evaluated on varied external test sets.



\section*{Introduction}
Since the emergence of the novel coronavirus disease (COVID-19) and the consequent global pandemic, healthcare authorities have used different diagnostic technologies to rapidly and accurately detect infected cases. Among such diagnostic technologies, chest Computed Tomography (CT) scans have been widely used, providing informative images of the lung parenchyma. More importantly, CT scans are highly sensitive to the diagnosis of COVID-19 infection, particularly based on its specific abnormality pattern and infection distribution in the lung~\cite{Fang2020}. To analyze a CT scan, radiologists should review several 2D images (slices), jointly creating a 3D representation of the body. Consequently, the analysis of a CT scan requires careful review of all slices. Furthermore, the COVID-19 lung imaging manifestations are highly overlapped with those of the Community Acquired Pneumonia (CAP), making the diagnosis even more challenging for radiologists.
The aforementioned issues have motivated the development of Artificial Intelligence (AI)-based diagnostic solutions using advancements in Deep Learning (DL) to analyze volumetric CT scans and provide diagnostic labels in a timely fashion~\cite{Mohammadi2021}. Despite the recent surge of interest and success of DL-based diagnostic solutions, such models commonly fail to achieve acceptable performances when there is heterogeneity in the data characteristics between the train and test sets, which is common when acquiring data from multiple imaging centers~\cite{Zhang2020}. Therefore, the necessity of developing a robust framework is of utmost importance to minimize the effect of the gap between the train and test sets and provide acceptable results on varied external datasets. In the case of CT scans, there are several factors contributing to the characteristics of the images among which, type of scanners, scanner manufacturers, and scanning protocols have the most influence on the quality and characteristics of the scans~\cite{Meyer2019,He2016}. Furthermore, the patients' clinical and surgical history can add more complexity and undesired artifacts to the CT scans that might have been blind to the trained model~\cite{Li2020}.

\begin{figure}[t!]
\centering
\includegraphics[width=0.9\linewidth]{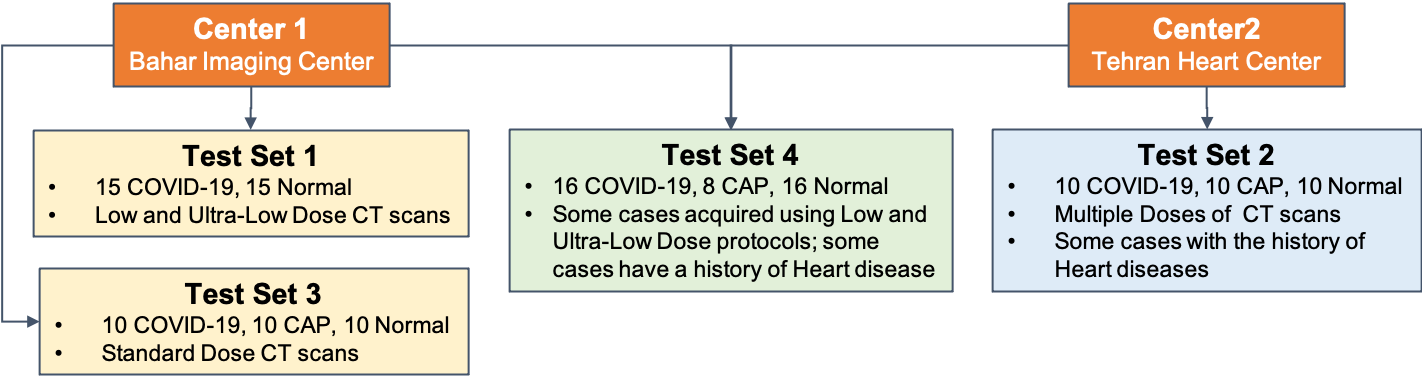}
\caption{\small Overview of the SPGC-COVID dataset used in this study.}
\label{fig:testset}
\end{figure}

Capitalizing on the above discussion, this study aims to develop a robust deep learning-based framework that can be generalized on varied external datasets with high flexibility to update itself upon receiving new external datasets. In this context, on the one hand, the paper introduces an automated two-stage classification framework based on Capsule Networks, which is tailored to robustly classify volumetric chest CT scans into one of the three target classes (COVID-19, CAP, or normal). The proposed Capsule Network-based framework integrates a scalable enhancement approach to boost its performance and robustness in the presence of gaps between the train and test sets regarding types of scanners, imaging protocols, and technical parameters. Furthermore,  this paper summarizes the 2021 Signal Processing Grand Challenge (SPGC) on COVID-19 diagnosis (SPGC-COVID challenge), which the authors organized as part of the 2021 IEEE  International Conference on Acoustics, Speech, and Signal Processing (ICASSP). In particular,  an overview of the top six models~\cite{Chaudhary2021, Yang2021, Garg2021, Xue2021, Bougourzi2021, Bingyang2021}  developed in the challenge is provided,  and their main components are investigated. In addition, the paper introduces a unique test dataset, referred to as the SPGC-COVID dataset, which is available for public access through Figshare~\cite{Heidarian2021b}. This dataset was used as the test set of the SPGC-COVID challenge. SPGC-COVID dataset consists of COVID-19, CAP, and normal cases acquired with various imaging settings from different medical centers. The SPGC-COVID dataset contains four subsets, illustrated in Fig.~\ref{fig:testset}, including images with different slice thickness, radiation dose, and noise level. In addition to different technical parameters, the dataset consists of CT scans of patients who have heart diseases or have undergone heart surgery, beside having COVID-19 or CAP infections. It is worth noting that the labels of the SPGC-COVID dataset were not released during the related competition, and participants had only access to the CT images,  not the labels. In this study, however, the associated labels are presented along with a comprehensive description of each test set, and a detailed list of technical parameters used to acquire the dataset. The performance of our proposed Capsule Network-based framework is compared with state-of-the-art approaches of the COVID-19 grand challenge. The results demonstrate that our proposed framework outperforms all the submitted models by achieving the overall accuracy of $96.15\%$ ($95\%CI: [91.25-98.74]$), COVID-19 sensitivity of $96.08\%$ ($95\%CI: [86.54-99.5]$), CAP sensitivity of $92.86\%$ ($95\%CI: [76.50-99.19]$), normal sensitivity of $98.04\%$ ($95\%CI: [89.55-99.95]$), and the Area Under the ROC curve (AUC) of $0.992$.

\section*{Materials and Methods}
\label{sec:methods}
In this section, first, the SPGC-COVID datasets is introduced. Then,  a brief summary of the 2021 SPGC-COVID challenge is provided, followed by the description of the challenge's best approaches.  Finally, the main components of our proposed two-stage Capsule Network-based classification framework are presented,  and the introduced unsupervised enhancement approach is described in detail.

\vspace{.1in}
\noindent
\textbf{Dataset}\\
\noindent
In what follows, different datasets used in this study are described individually, followed by supplementary information about the demographic data, imaging protocols, acquisition settings, de-identification, and the labeling process. The utilized dataset consists of a training and a test set, where the training dataset is the COVID-CT-MD~\cite{Afshar2021a}, we introduced previously and is acquired from one imaging center using similar scanning parameters. The test dataset, the so-called SPGC-COVID, is comprised of four different sets each with specific characteristics to evaluate the robustness and generalizability of the DL model from different aspects.

\begin{figure}[t!]
\centering
\includegraphics[width=0.5\linewidth]{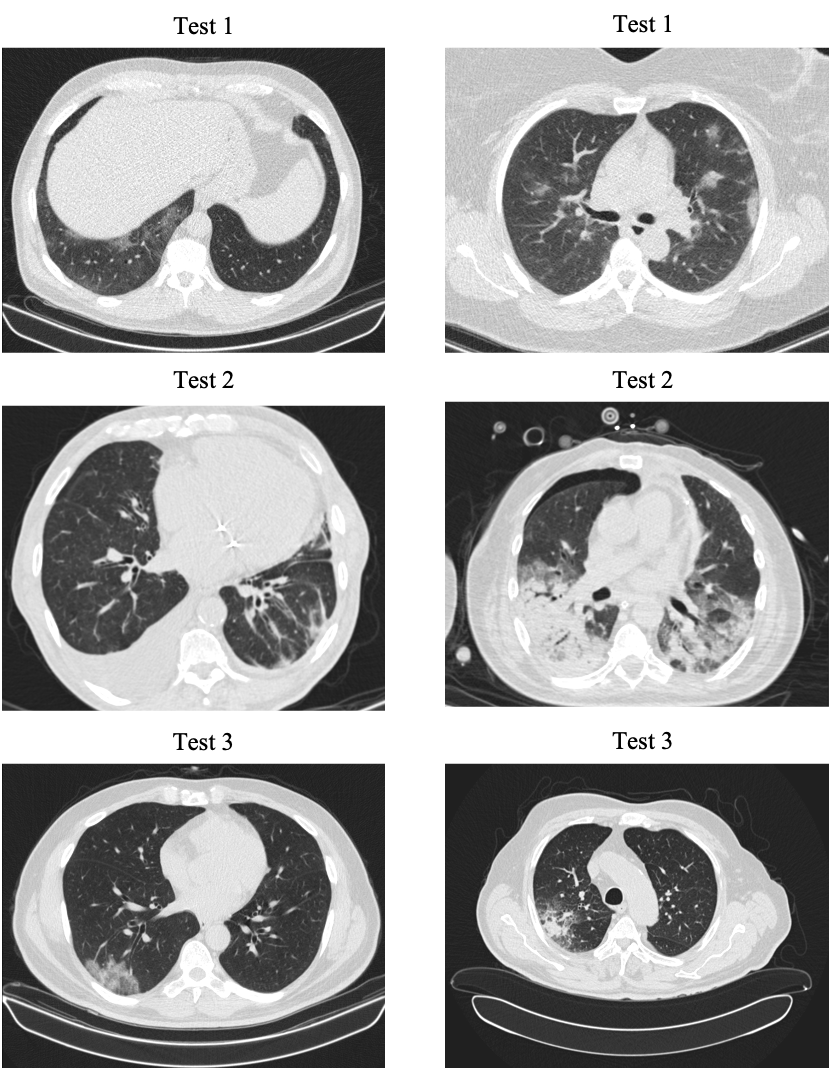}
\caption{\small Sample CT slices from the first three test sets. In Test set 1, the noise level is high. In Test 2, some cases reveal cardiovascular-related complications. In Test 3, the image quality and contrast are higher compared to other test sets.}
\label{fig:sample-ct}
\end{figure}

An overview of different datasets and imaging centers is visualized in Fig.~\ref{fig:testset}. Different components of the utilized dataset are as follows:
\begin{itemize}[noitemsep]
\item \textbf{Train Set:} We used our in-house and publicly available dataset~\cite{Afshar2021a}, referred to as the ``COVID-CT-MD", as the training dataset which contains CT scans of COVID-19, CAP, and normal cases acquired by the ``SIEMENS, SOMATOM Scope" scanner using the standard radiation dose from Babak Imaging Center, Tehran, Iran. A subset of $55$ COVID-19, and $25$ CAP cases are analyzed by one radiologist (M.J.R.) to identify slices demonstrating infection. The labeled subset of the data contains $4,993$ slices demonstrating infection and $18,416$ slices without evidence of infection. $30\%$ of the cases in this set are randomly selected as the validation set.
\item \textbf{The SPGC-COVID Test Set:} This dataset, which is released through this manuscript, comprises the following four different subsets:
\begin{itemize}[noitemsep]
\item \textbf{Test Set 1:} Low and Ultra-Low dose CT scans of COVID-19 and normal cases acquired from the same imaging center as that of the train set. This dataset is a subset of our in-house dataset of Low Dose CT scans~\cite{Afshar2021b} and is publicly available.
\item \textbf{Test Set 2:} CT scans of COVID-19, CAP, and normal cases acquired in a different imaging center (Tehran Heart Center, Iran) using the ``SIEMENS SOMATOM Emotion 16" scanner and different scanning parameters. Some cases in this dataset have additional history of cardiovascular disease/surgeries with specific CT imaging findings, which are not available in the train set.
\item \textbf{Test Set 3:} CT scans of COVID-19, CAP, and normal cases obtained by the same scanner and scanning protocol used to acquire the train set. Cases in this test set are not included in the COVID-CT-MD public dataset.
\item \textbf{Test Set 4:} A combination of new CT scans of all three categories (i.e.,  COVID-19, CAP, Normal) obtained from the same centers as those of Test set 1 and 2, using the same acquisition settings and scanners.
\end{itemize}
\end{itemize}
%
Additional statistical and demographic information about different train and test sets used in this study are provided in Table~\ref{tab:dataset-detail}. In Table~\ref{tab:dataset-detail}, Center 1 represents the Babak Imaging Center and Center 2 is the Tehran Heart Center. Both imaging centers are located in Tehran, Iran and use the Filtered Back Projection reconstruction method~\cite{Pontana2011} to obtain the CT images. Sample CT slices from the first three test sets are shown in Fig~\ref{fig:sample-ct}. Various scanning protocols and settings have been used to obtain the train and test datasets used in this study. The important parameters that contribute the most to the image quality and characteristics are presented in Table~\ref{tab:scan-protocol}.

\begin{table}[t!]
\begin{adjustwidth}{-2.25in}{0in}
\centering
\begin{tabular}{|c|c|c|c|c|c|}
\hline
Dataset  & Slice Thickness (mm) & Reference Exposure (mAs) & kVp (kV) & Radiation (mSv) & \makecell{Number of Slices\\(per patient)}\\
\hline
Train & $2$ & $50$ & $110-130$ & $\sim7$ & $68-195$ \\
\hline
Test 1  & $2$ & $15-20$ & 110 & $\sim0.3-1.5$ & $126-169$ \\
\hline
Test 2& $1.5-5$ & $25$ & $110-130$ & $\sim2$ & $53-221$ \\
\hline
Test 3 & $2$ & $50$ & $100-110$ & $\sim7$ & $115-183$ \\
\hline
Test 4 & $1.5-6$ & $15-25$ & $110-130$ & $\sim0.3-2$ & $52-224$ \\
\hline
\end{tabular}
\caption{\small Acquisition parameters used to obtain each dataset.}
\label{tab:scan-protocol}
\end{adjustwidth}
\end{table}

\vspace{.1in}
\noindent
\textbf{\textit{De-identification:}} The data used in this study complies with the DICOM supplement 142 (Clinical Trial De-identification Profiles)~\cite{Committee2011}, which ensures that all personal information is removed or obfuscated, including names, UIDs, dates, times, comments, and center-related information. Some demographic and acquisition attributes related to the patients' gender and age, scanner type, and image acquisition settings have been preserved to provide useful information about the dataset.

\vspace{.1in}
\noindent
\textbf{\textit{Labeling Process:}} Diagnosis of the cases scanned in Center 1 is obtained by finding the consensus between three experienced radiologists who have considered the following three main criteria to label the data:
\begin{enumerate}[noitemsep]
\item[(i)] RT-PCR test (if available);
\item[(ii)] Imaging findings including Ground Glass Opacities (GGOs), consolidations, crazy paving pattern, bilateral and multifocal lung involvement, peripheral distribution, and lower lobe predominance of findings;
\item[(iii)] Clinical symptoms of the COVID-19 infection, and;
\item[(iv)] Epidemiology.
\end{enumerate}
For the cases acquired from Center 2, $(13/18)$ COVID-19 cases have positive RT-PCR test results and the remaining cases have been labeled by one experienced radiologist following the same aforementioned criteria. Slice-level labels are provided by one radiologist to identify and label slices with evidence of infection. A subset of $15$ random cases has been further reviewed by the two other radiologists to confirm the accuracy of the slice-level labels.

\vspace{.1in}
\noindent
\textbf{Summary of the 2021 SPGC-COVID Challenge}\\
\noindent
In the first phase of this SPGC-COVID challenge,  participants had access to the same train and validation sets as those used in this study to develop and evaluate their models. In the second phase, they have been provided with the first three test sets and had two weeks to submit their final models. Finally, the best-performing models based on the first three test sets have been evaluated on the fourth test set to determine the overall performances. In what follows, the main components of the six best-performing models in the SPGC-COVID challenge~\cite{Chaudhary2021, Yang2021, Garg2021, Xue2021, Bougourzi2021, Bingyang2021}  are briefly described.  As stated previously, an external model~\cite{Nicolas2021} developed based on the same dataset,  but not as part of the challenge,  is also investigated and used for comparison in this study.  This model is also summarized below.
\begin{itemize}
\item \textbf{Ref.~\cite{Chaudhary2021}:} In this model, slice-level predictions are acquired from an EfficientNet-based classifier~\cite{Tan2019} and a weighted majority voting is proposed to obtain the final patient-level labels. To train this classifier, the authors first trained two separate binary classifiers to detect slices demonstrating infection from COVID-19 and CAP cases. Then, they fed these models with unlabelled cases to provide the training set for the main classifier. Additionally, they only considered the middle slices (e.g., 80 middle slices) in a volumetric CT scan at the training phase.
\item \textbf{Ref.~\cite{Yang2021}:} This model aggregates the output of six classifiers developed based on the 3D ResNet101 model~\cite{He2015}. One model in this proposed framework is a three-way classifier trained over all of the cases while the other five models are binary classifiers independently trained over COVID-19 and CAP cases using different combinations of train and validation sets.
\item \textbf{Ref.~\cite{Garg2021}:} This model presents a feature extraction-based approach in which a modified pre-trained ResNet50 model classifies each slice into the target classes and the penultimate fully connected layer is extracted as the feature map. Next, a max-pooling layer followed by two fully connected layers are used to generate patient-level prediction from slice-level feature maps. The output of this model is then aggregated with two BiLSTM patient-level classifiers, which are fed by the same slice-level feature maps to provide the final patient-level labels.
\item \textbf{Ref.~\cite{Xue2021}:} The pre-trained 3D Resnet50~\cite{Ebrahimi2020} is the backbone of this model. The authors first doubled the number of slices for each case using a 3D cubic interpolation method. Then, they extracted the lung area using a pixel-based segmentation approach, followed by classical image processing techniques such as pixel filling and border cleaning. Finally, a subset of slices is selected from each volumetric CT scan based on their lung area and an experimentally-set threshold, which are then resized into a ($224,224,224$) data, using a 3D cubic interpolation method, providing the patient-level input for training and evaluation purposes.
\item \textbf{Ref.~\cite{Bougourzi2021}:} This model utilizes a two-stage framework in which the first stage is responsible for performing a multi-task classification to classify 2D slices into one of the target groups and identify the location of the slice in the sequence of CT images at the same time. The model at the first stage uses an ensemble of four popular CNN-based classifiers (i.e., ResneXt50~\cite{Xie2017}, DenseNet161~\cite{Huang2017}, Inception-V3~\cite{Szegedy2016}, and Wide-Resnet~\cite{Zagoruyko2016}), followed by an aggregation mechanism that divides the whole volumetric CT scan into 20 groups of slices and calculates the percentage of infected slices related to COVID-19 and CAP classes in each group. The values obtained for all groups are then concatenated and fed into a XG-boost classifier~\cite{Chen2016} in the second stage to generate patient-level predictions.
\item \textbf{Ref.~\cite{Bingyang2021}:} The model proposed in this work initiates with a slice-level EfficientNet-B1 classifier~\cite{Tan2019} aiming to classify slices and generate feature maps (intermediate layers) to be used in the subsequent sequence classifier. In the sequence classifier, several weak classifiers are trained and the outputs are aggregated using an adaptive weighting mechanism to obtain the final patient-level results. To further enhance the performance of the model and cope with the imbalanced training set, a combination of weak and strong data augmentations are applied to the training cases, forcing the model to produce similar labels for both types of augmented images. Furthermore, to improve the robustness of the model when being tested on varied datasets, a K-Means clustering method ($K=3$)~\cite{macqueen1967} is adopted to develop a single classifier for each cluster of the data and aggregate the results via a majority voting approach.
\end{itemize}
The following provides a brief description of the model, which is not proposed in the SPGC-COVID challenge, but is used for comparison in this study as similar datasets are used for  development and evaluation of the model:
\begin{itemize}
\item \textbf{Ref.~\cite{Nicolas2021}:} This model aims to introduce a robust training algorithm and classification framework, which is capable of being updated upon receiving new datasets to deal with the characteristic shifts in different test sets. First, it adopts a two-stage architecture similar to the COVID-FACT model proposed in reference~\cite{Heidarian2021} and trains the benchmark model in a self-supervised fashion~\cite{Jing2020} and the majority voting is adopted to obtain patient-level labels. The backbone model used in this study is DenseNet169~\cite{Huang2017} and strict slice preprocessing and sampling methods are applied to the training set. Such methods contain pixel-based approaches with some fixed thresholds used to extract lung areas and select the slices with the most visible lung area. Next, each test set is divided into four quarters, which are then used in an unsupervised updating process, in which quarters are passed to the model sequentially and confident predictions are selected to fine-tune the slice-level classifiers. A slice-level prediction is considered confident in this study if it achieves the probability of at least $0.9$ in agreement with the patient-level label. 
\end{itemize}
The key components of our proposed model and those used for comparison are summarized in Table~\ref{tab:models-details}. All models resized and normalized the input data to be compatible with the utilized architectures. In addition, most frameworks (except those using a 3D model as their backbone) adopted a multi-stage framework transferring information from the slice-level domain to the patient-level one, some of which also utilized an ensemble architecture to aggregate the extracted information.

\vspace{.1in}
\noindent
\textbf{Proposed Capsule Network-based framework}\\
\noindent
In this study, we have developed a two-stage framework similar to the model proposed in our previous study~\cite{Heidarian2021}, referred to as the ``COVID-FACT", as our benchmark model to classify volumetric CT scans into three target classes of COVID-19, CAP, and normal. We then use the unlabeled data from the test sets to boost the performance and robustness of the framework on the unseen cases. The pipeline of the proposed framework is shown in Fig.~\ref{fig:pipeline}.

It is worth mentioning that although Capsule Network is the building block of models proposed in this study and our previous study~\cite{Heidarian2021}, these two frameworks target different challenges, and different experiments are performed in the associated studies.
The main contributions of this study and its key differences with the previous study are outlined as follow:
\begin{enumerate}
\item This study aims to distinguish COVID-19, CAP, and normal cases (a three-way classification is considered), whereas in the case of the COVID-FACT framework, only COVID and non-COVID cases are classified and a binary classification is considered.  It is worth noting that due to similarities between COVID-19 and CAP cases, the three way classification is more challenging and previously developed models cannot be applied directly.
\item This study aims to address the problem of generalizeability of deep learning-based models, which has been debated for a long time. The COVID-FACT framework, however, was developed soon after the emergence of the pandemic and collecting multi-center datasets was not possible at the time. In this study, the aforementioned problem is further investigated to take one step forward towards reaching a clinically applicable AI-based approach.
\item A scalable unsupervised enhancement mechanism is proposed to increase the model's performance upon receiving new datasets from different centers and cohorts.
\item In this study, several modifications have been applied to different stages of the COVID-FACT framework, including adding residual connections and dropout layers to allow the model to capture more informative features and overcome the over-fitting problem.
\item The SPGC-COVID dataset is introduced, which contains four different subsets each with specific characteristics in terms of scanning setting, imaging center, and clinical background of the subjects. The 2021 SPGC-COVID challenge is summarized and the best-performing models submitted to the challenge are described and investigated. 
\end{enumerate}

\begin{landscape}
\begin{table}
\begin{center}
{\color{black}
\begin{tabular}{ |c|c|c|c|c|c|c|c|c| }
 \hline
  \textbf{Framework} & \textbf{\makecell{Lung\\Extraction}} & \textbf{\makecell{Imbalanced\\Data\\Mitigation}} & \textbf{Backbone(s)}& \textbf{\makecell{Aggregation}} & \textbf{\makecell{Heterogeneity\\Consideration}} & \textbf{\makecell{Thresholds \&\\Coefficients}} & \textbf{pre-Training} & \textbf{\makecell{Data\\Aug.}} \\
  \hline
 \textbf{Proposed} & \makecell{U-Net R231\\COVIDweb} & \makecell{Weighted\\loss function} & CapsNets & \makecell{MV} & \makecell{Unsupervised\\update\\based on\\confident\\predictions} & 2 & None & \xmark  \\
  \hline
 \textbf{\makecell{Ref.~\cite{Chaudhary2021}}} &  None & None & \makecell{DenseNet-121\\Efficient Net} & \makecell{WMV \&\\GAP} & None
 & 3 & ImageNet & \xmark \\
 \hline
  \textbf{\makecell{Ref.~\cite{Yang2021}}} &  \makecell{U-Net R231\\COVIDweb} & \makecell{Different\\balanced\\training\\samples} & 3D ResNet101 & None & None & 0 & None & \cmark \\
 \hline
  \textbf{\makecell{Ref.~\cite{Garg2021}}} & Image Contours & None & ResNet50 & GAP  & None & 1 &  COVIDx-CT & \cmark \\
 \hline
  \textbf{\makecell{Ref.~\cite{Xue2021}}} & \makecell{Pixel\\intensity} & \makecell{Up-sampling\\minority class}& 3D Res	Net50 & None & None & 4 & ImageNet & \cmark \\
 \hline
  \textbf{\makecell{Ref.~\cite{Bougourzi2021}}} & \makecell{Custom\\Encoder-Decoder} & \makecell{Down-sampling\&\\Up-sampling} & \makecell{ResNet50\\DenseNet161\\Inception-V3\\XG-boost\\Wide-ResNet} & \makecell{Feature\\concatenation} & None & 0 & ImageNet & \cmark  \\
 \hline
  \textbf{\makecell{Ref.~\cite{Bingyang2021}}} & None & None & \makecell{AdaBoost\\EfficientNet} & WV & \makecell{Adaptive\\Boosting\&\\Data\\Clustering} & 1 & ImageNet & \cmark \\
 \hline
  \textbf{\makecell{Ref.~\cite{Nicolas2021}}} & None & \makecell{Cut-off\\threshold\\adjustment} & DenseNet169 & GAP & \makecell{Online\\unsupervised\\learning} & 2 & \makecell{Self-supervision\\on the same\\dataset} & \cmark \\
 \hline
\end{tabular}}
\end{center}
\caption{Underlying key features of the proposed framework and seven models used for comparison in this study (i.e., top six models developed following the SPGC-COVID challenge, and one model developed outside of the scope of this challenge).
\\
MV: Majority Voting, WMV: Weighted Majority Voting, WV: Weighted Voting, GAP: Global Average Pooling, NA: Not Available}
\label{tab:models-details}

\end{table}
\end{landscape}

\begin{figure}[t!]
\centering
\includegraphics[width=1\linewidth]{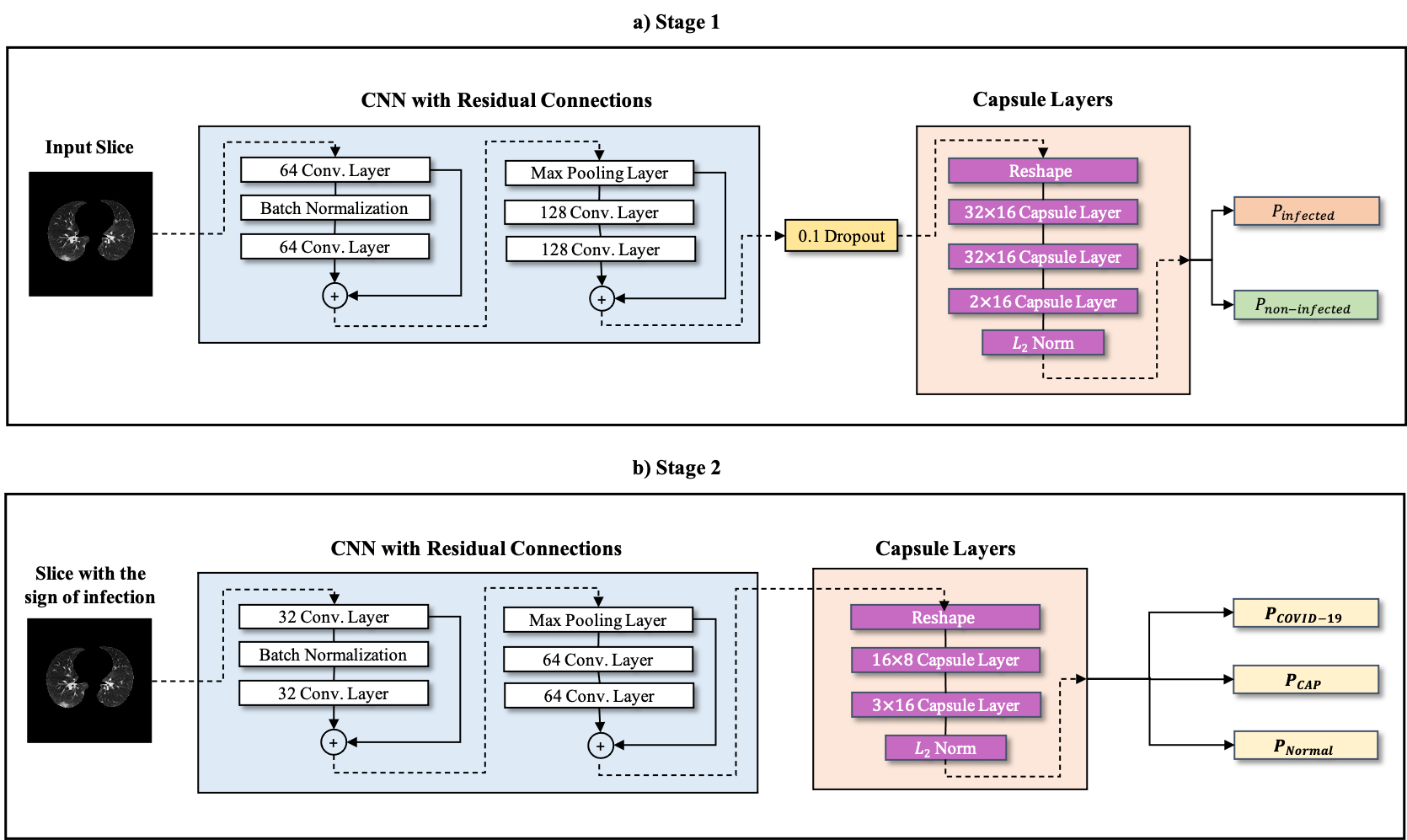}
\caption{\small a) The structure of the CapsNet binary classifier in stage 1. b) The structure of the three-way classifier in stage 2. $+$ sign denotes the residual addition.}
\label{fig:stage1}
\end{figure}

In what follows, different components of the proposed framework are described:
\begin{itemize}
\item\textbf{Preprocessing:} Raw CT scans, typically, contain uninformative components and unwanted artifacts (e.g., metallic artifacts), which can negatively affect the performance of the DL model. In addition, image sizes may vary and pixel intensities may be in different ranges when the images are acquired by different scanners. As such, we first extracted the lung areas from the CT images to remove the insignificant and distracting components. In this regard, we used a well-trained U-net based segmentation model~\cite{Hofmanninger2020}, which is fine-tuned on COVID-19 cases to specify lung areas in the first step. We then down-sampled all images into the ($256\times256$) size to reduce the memory allocation and complexity without significant loss of information. Furthermore, we normalized each 2D image into the $[0,1]$ interval.
\item\textbf{Stage 1:} The first stage performs the infection identification task, which aims to find slices with the evidence of infection (caused by CAP or COVID-19) for each patient. The identified slices will then be classified into one of the three target classes in the second stage. The input of Stage 1 is the normalized lung area as a 2D image and the output is the label indicating whether the input image demonstrates infection or not. The classification model used in this stage is based on the Capsule Networks (CapsNets)~\cite{hinton2018}, which have shown a superior discriminative capability compared to their CNN-based counterparts, especially when they are trained over small datasets~\cite{Afshar2021, Afshar2020, Heidarian2021a, Afshar2021c}. Each capsule layer consists of multiple capsules, which are groups of neurons represented by a vector. Capsule Network benefits from an iterative process, known as the ``Routing by Agreement", that aims to evaluate the agreement between the capsules in a lower layer on the existence of an object in the higher layer. Using the Routing by Agreement process, the model can recognize the relation between multiple instances in an image. Furthermore, CapsNets have lower time and space complexity compared to the conventional CNNs~\cite{Heidarian2021}. Such advantages make CapsNet-based models ideal in the case of COVID-19 where small annotated datasets are available and disease manifestations show specific spatial distributions in the lung. The detailed structure of the classification model in the first stage is shown in Fig.~\ref{fig:stage1}(a).\\
For the first stage, we adopted the same architecture as the model proposed in~\cite{Heidarian2021}. More specifically, the model in this stage uses a stack of four convolution layers, one batch normalization layer, and one max pooling layer to generate initial feature maps. Next, the output of the last convolution layer is reshaped to form the first capsule layer, followed by three consecutive capsule layers, as shown in Fig.~\ref{fig:stage1}(a). The last layer contains two capsules representing the two target classes (i.e., slices with and without the evidence of infection.) The length of each capsule represents the probability of the corresponding class being present. Different from COVID-FACT, residual connections are added between the convolution layers to transfer low-level features to the deeper layers. This modification further assists the model in identifying informative features. Additionally, we have added a dropout layer before the capsule layers to overcome the overfitting problems during the training. The labeled subset of the training dataset has been used to train this stage over $100$ epochs using the Adam optimizer with the learning rate of $1e-4$. To account for the imbalanced number of slices in each class, we have used a weighted loss function to increase the contribution of the minority group (i.e., slices demonstrating infection) to the final loss value and balance the influence of each class. The balanced loss function used to train Stage 1 is given by
\begin{equation}
\begin{split}
loss & = w_1\times loss_1 + w_2 \times loss_2,\\
& w_1 = \frac{N_2}{N_1 + N_2},\\
& w_2 = \frac{N_1}{N_1 + N_2},
\end{split}
\label{eq:loss}
\end{equation}
where $w_1$ and $w_2$ represent the weights corresponding to the loss value calculated for negative and positive samples, respectively. Term $loss_1$ denotes the loss associated with negative samples, while $loss_2$ is the loss associated with positive samples. Term $N_1$ represents the number of negative samples, and $N_2$ is the number of positive samples.
\item\textbf{Stage 2:} The second stage takes the candidate slices from the previous stage and classifies them into one of the COVID-19, CAP, or normal cases. More specifically, we have used the slices demonstrating infection recognized by the first stage for all of the cases in the train set (with or without slice level labels) to train a three-way classification model.
Stage 2 utilizes a CapsNet architecture similar to the one used in the first stage but with smaller dimensions and three capsules in the last layer to represent three target classes. The architecture of stage two is shown in Fig.~\ref{fig:stage1}(b). Similar to the first stage, we used a weighted loss function to cope with the imbalanced number of samples in some categories. At this stage, the loss weights associated with normal and CAP classes are set to $5$ and the weight for COVID-19 class is set to $1$. Note that as the normal cases are extremely rare at this stage, the weights are set differently compared to those calculated by Eq.~\ref{eq:loss}, to maintain the stability of the training process, while enforcing the model to pay more attention to the minority classes. We also used the binary cross-entropy loss function, which translates the three-way classification problem at hand into three binary classification tasks. In fact, the loss value is calculated separately for each binary label associated with a target class (i.e., COVID-19, CAP, normal). Finally, a majority voting mechanism is adopted to transfer slice-level predictions into patient-level ones and determine the final label. It is worth noting that an accurate model in the first stage detects only a few candidate slices from normal cases. We can then apply a thresholding mechanism on the output of the first stage to identify those cases with only a few identified infectious slices in the first stage and label them as normal. We have used a threshold of $3\%$ to specify normal cases immediately after the first stage. More specifically, if less than $3\%$ of the slices in a volumetric CT scan are classified as infectious, the corresponding CT scan is classified as a normal case. Based on~\cite{Yu2020}, the minimum lung lesion involvement in patients with COVID-19-related CT findings is $4\%$. In addition, the minimum percentage of slices demonstrating infection in our training dataset is $7\%$. In the case that the model in stage 1,  misclassifies more than $3\%$ of slices for a normal case, there is still a chance to classify the slices as normal in the second stage.
\end{itemize}
\noindent
\textbf{\textit{Unsupervised Enhancement:}} Unseen CT scans acquired by different scanners and scanning protocols contain heterogeneous characteristics leading to lower performance of a pre-trained model. To increase the robustness, we take advantage of the extra unlabeled samples that are available via the various test cases, and utilize this extra set of CT scans in an unsupervised fashion. In other words, inspired by the ideas from ``Active Learning~\cite{Smailagic2018,Smailagic2020, Budd2021}'', where different data samples are extracted to train the model in different stages, and ``Semi-Supervised Learning~\cite{cai2013, Schmarje2021}'', where a label is assigned to unlabelled cases based on a pre-defined metric, we developed an autonomous mechanism to extract and label a part of data in the test sets using a probabilistic selection criteria with reduced complexity. The selected sample and the assigned labels are then used to re-train and boost the initially trained model. More specifically, we selected those test cases for which the model generated the most confident results (i.e., high probability). Similarly, among the selected cases, those with high confidence in slice-level predictions are used. To define the confident results,  the probability of a volumetric CT scan belonging to a specific target class is considered to be equal to the ratio of the slices belonging to that class over the total number of slices (all slices containing the lung lesion), which can be written as follows
\begin{equation}
P(\mathbb{X} \in \mathcal{C}_i) = \frac{n_{\mathcal{C}_i}}{\sum_{i=1}^{C}{n_{\mathcal{C}_i}}},
\end{equation}
where $\mathbb{X} $ represents the input volumetric CT scan, $C$ represents the number of target classes, and $n_{\mathcal{C}_i}$ denotes the number of slices belonging to the target class $\mathcal{C}_i$. Then, we introduced a confidence threshold value and considered a prediction confident if the probability of the input CT scan belonging to any of the target classes is more than the pre-set threshold. In this study, we have used $80\%$ as the confidence threshold. A similar approach is used to extract confident slices and their corresponding labels. In this case, the probability of a slice belonging to a target class is determined by the output of the CapsNet classifier in Stage 2, which is the length ($L_2 Norm$) of capsules in the last layer. It is worth mentioning that for those normal cases, which are identified in the first stage using the described thresholding mechanism, we only select the slices which are misclassified as infectious with a high probability (e.g., more than the confidence threshold). Such slices will be labeled as normal in the enhancement phase.
Following the aforementioned steps, we can obtain a set of slices and their corresponding labels to augment the training dataset aiming to make the model more aware of the new features available in the unseen datasets and achieving more robust feature maps. Therefore, for each test set, we obtained a set of confident slices and their associated labels which have been added to the train set to re-train the model of the second stage. It is worth noting that the first stage has been kept unchanged in this approach. Finally, after re-training the benchmark model based on the confident slices acquired from each test set, we have obtained several enhanced models (each related to one test set) and averaged the associated patient-level probability scores to achieve the final prediction.
This aggregation mechanism depends on the target test set. More specifically, to apply the model on each test set, we take the average of the predictions obtained by the models enhanced over the other test sets. For instance, the model developed for the diagnosis of cases in Test set 1 takes the average of probability scores provided by the models enhanced on Test set 2 and 3. The main reason for using such an aggregation mechanism is that the enhancement based on a specific test set will further boost the probability scores of confidently predicted slices while having limited influence on other cases in the same set. As such, incorporating the model enhanced on a test set will not bring in any further improvement to the evaluation process of the same set. The results presented in Table~\ref{tab:ind-result} further support this discussion. It is worth noting that we used the first three test sets to enhance the benchmark model and kept the fourth test set aside for only evaluation purposes. As such, upon receiving new test datasets, we can aggregate the results of the enhanced models on the individual test sets (each representing a specific center or scanning protocol) to provide the classification results for the new cases. The unsupervised model enhancement described above along with the subsequent ensemble averaging make the entire framework a robust automated framework that can be easily improved and updated upon receiving new datasets from different imaging centers.

\noindent
\section*{Results}
\begin{figure}[t!]
\centering
\includegraphics[width=1\linewidth]{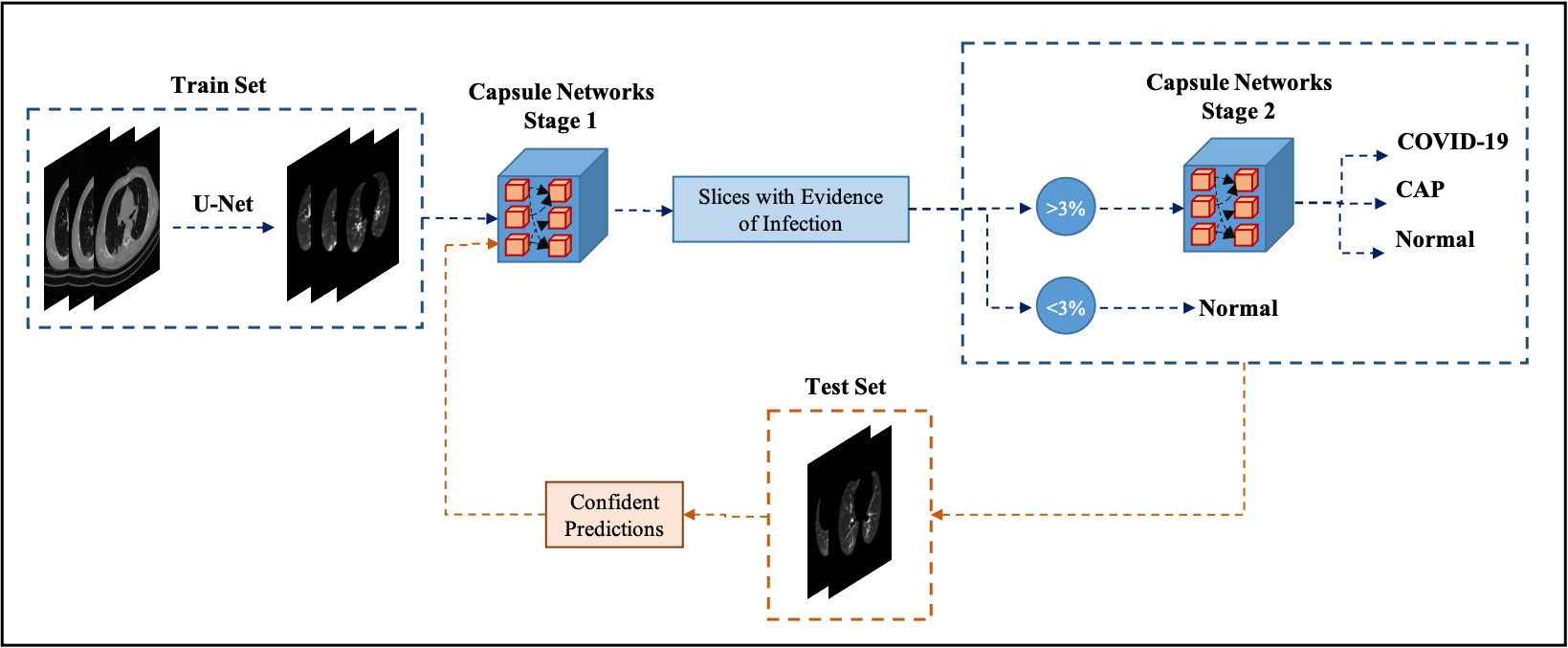}
\caption{\small The pipeline of the proposed CapsNets-based framework.}
\label{fig:pipeline}
\end{figure}

Our proposed framework adopts a two-stage architecture based on Capsule Networks (CapsNets)~\cite{hinton2018}, as shown in Fig.~\ref{fig:pipeline}, which is fed by a volumetric CT scan and provides the probability of the input scan belonging to one of the three target classes. In brief, the first stage identifies CT slices demonstrating infection and passes them to the second stage to be classified as one of the target classes. The output of the first stage is also used to filter normal cases, by applying a $3\%$ threshold on the involvement of the lung parenchyma (i.e., the ratio of the infectious slices in the whole volume). In addition to the proposed framework, four partially enhanced models are developed (based on the four test sets), and the final model aggregates the outputs of the partially enhanced models to provide the final predictions. The proposed enhancement approach extracts confidently predicted images from each test set in an unsupervised fashion, which are then used to update the model's parameters.

\begin{table}[t!]
\begin{adjustwidth}{-2.25in}{0in}
\centering
\begin{tabular}{|c|c|c|c|c|c|c|}
\hline
\textbf{Dataset} & \textbf{COVID-19} & \textbf{CAP} & \textbf{Normal} & \textbf{Age (Mean$\pm$ SD)} & \textbf{Gender} & \textbf{Imaging Center}\\
\hline
Train & 171 & 60 & 76 & $50.78\pm16.84$ & $183M/124F$ & 1\\
\hline
Test 1 & 15 & 0 & 15 & $40.97\pm14.38$ & $19M/11F$ & 1 \\
\hline
Test 2 & 10 & 10 & 10 & $61.00\pm13.39$ & $25M/5F$& 2 \\
\hline
Test 3 & 10 & 10 & 10 & $46.77\pm20.89$ & $15M/15F$& 1 \\
\hline
Test 4 & 16 & 8 & 16 & $46.23\pm14.74$ & $25M/15F$ & Both \\
\hline
\end{tabular}
\caption{\small Number of cases, demographic data, and additional acquisition information for each dataset. Center 1 represents the Babak Imaging Center and Center 2 is the Tehran Heart Center. Both imaging centers are located in Tehran, Iran.}
\label{tab:dataset-detail}
\end{adjustwidth}
\end{table}

To evaluate the performance of the proposed model and the effectiveness of its unsupervised enhancement approach, we used the first three test sets to enhance the benchmark model and kept the fourth test set aside only for evaluation purposes. The results obtained by applying the enhanced ensemble model on all of the test sets are shown in Table~\ref{tab:result}. In this table, the Area Under the ROC curve (AUC) is calculated based on the micro average of the values obtained for each class. In addition, to further validate the obtained results, confidence intervals for the total accuracy and sensitivity are provided using the method introduced in~\cite{Agresti1998}.

\begin{table}[t!]
\begin{adjustwidth}{-2.25in}{0in}
\centering
\begin{tabular}{|c|c|c|c|c|c|}
\hline
\textbf{Test Set}& \textbf{Accuracy(\%)} & \textbf{\makecell{COVID-19\\Sensitivity(\%)}} & \textbf{\makecell{CAP\\Sensitivity(\%)}} & \textbf{\makecell{Normal\\Sensitivity(\%)}} & \textbf{\makecell{AUC\\ (micro)}}\\
\hline
Test 1& 100 & 100 & NA & 100 & 1.000\\
\hline
Test 2 & 86.67 & 80 & 90 & 90 & 0.952\\
\hline
Test 3 & 100 & 100 & 100 & 100 & 1.000\\
\hline
Test 4 & 97.50 & 100 & 87.50 & 100 & 0.999\\
\hline
Total & 96.15(CI: [91.25-98.74]) & 96.08(CI: [86.54-99.5]) & 92.86(CI: [76.50-99.19]) & 98.04(CI: [89.55-99.95]) & 0.992\\
\hline
\end{tabular}
\caption{Result obtained by the proposed framework for different test sets. $95\%$ Confidence Intervals obtained for total performance using the significance level of $0.05$ are presented in parentheses.}
\label{tab:result}
\end{adjustwidth}
\end{table}

To elaborate the effect of the proposed unsupervised enhancement approach, we have provided the performance of the benchmark model (i.e., before enhancement) as well as the models enhanced by individual tests sets (i.e., before averaging the outputs) in Table~\ref{tab:ind-result}. Results shown in Table~\ref{tab:ind-result} imply that the probability of the input CT scan belonging to the target class in some misclassified cases have been on the thresholding edge (close to $0.5$) and could be corrected after incorporating the models enhanced over other test sets.

\begin{table}[t!]
\begin{adjustwidth}{-2.25in}{0in}
\centering
\begin{tabular}{ | m{4em} | m{5.2em} | m{5.2em}| m{5.2em} |m{5.2em}|m{5.2em}|m{5.2em}| }
\hline
Test Set & Sensitivity & Proposed & Enhanced \#1 & Enhanced \#2 & Enhanced \#3 & Benchmark \\
\hline
Test1 &COVID-19 \hspace{35pt} Normal\hfill & \textbf{15/15} \hspace{35pt}\textbf{15/15} & 15/15 \hspace{35pt}15/15 & 15/15 \hspace{35pt}15/15 & 15/15 \hspace{35pt}15/15 & 15/15 \hspace{35pt}15/15 \\
\hline
Test2 &COVID-19 \hspace{35pt} CAP \hspace{35pt} Normal\hfill & \textbf{8/10} \hspace{35pt}\textbf{9/10} \hspace{35pt}\textbf{9/10} & 8/10\hspace{35pt} 9/10 \hspace{35pt}9/10 & 8/10 \hspace{35pt}8/10 \hspace{35pt}9/10 & 8/10 \hspace{35pt}9/10 \hspace{35pt}9/10 & 8/10 \hspace{35pt}8/10 \hspace{35pt}9/10 \\
\hline
Test3 &COVID-19 \hspace{35pt} CAP \hspace{35pt} Normal\hfill & \textbf{10/10} \hspace{35pt}\textbf{10/10} \hspace{35pt}\textbf{10/10} & 10/10\hspace{35pt} 10/10 \hspace{35pt}10/10 & 9/10 \hspace{35pt}10/10 \hspace{35pt}10/10 & 9/10 \hspace{35pt}10/10 \hspace{35pt}10/10 & 9/10 \hspace{35pt}10/10 \hspace{35pt}10/10 \\
\hline
Test4 &COVID-19 \hspace{35pt} CAP \hspace{35pt} Normal\hfill & \textbf{16/16} \hspace{35pt}7/8 \hspace{35pt}\textbf{16/16} & 16/16\hspace{35pt} 6/8 \hspace{35pt}16/16 & 16/16 \hspace{35pt}7/8 \hspace{35pt}16/16 & 15/16 \hspace{35pt}\textbf{8/8} \hspace{35pt}16/16 & 15/16 \hspace{35pt}7/8 \hspace{35pt}16/16 \\
\hline
\end{tabular}
\caption{The ratio of correctly classified cases over total cases in the associated class obtained for the proposed model, the benchmark model, and the partially enhanced models.}
\label{tab:ind-result}
\end{adjustwidth}
\end{table}

In addition to the final patient-level predictions, we have evaluated the performance of the first stage on the validation set in detecting slices demonstrating infection to have a clearer insight into the internal components of the framework. The first stage achieved an accuracy of $93.41\%$, sensitivity of $91.04\%$, and specificity of $94.26\%$ in the binary (infectious \& non-infectious) classification task. As slice-level labels (i.e., binary labels indicating the existence of infection in a CT slice) are not available for test sets, the result on the validation set is only reported. Moreover, as mentioned earlier, the output of the first stage can be used to identify most normal cases before entering the next stage. We found that nearly all of the normal cases in the four test sets ($45/46$ cases) have been identified correctly by the thresholding mechanism applied on the output of the first stage, while none of COVID-19 and CAP cases have been misclassified as normal using this thresholding approach.

\begin{figure}[t!]
\centering
\includegraphics[width=0.55\linewidth]{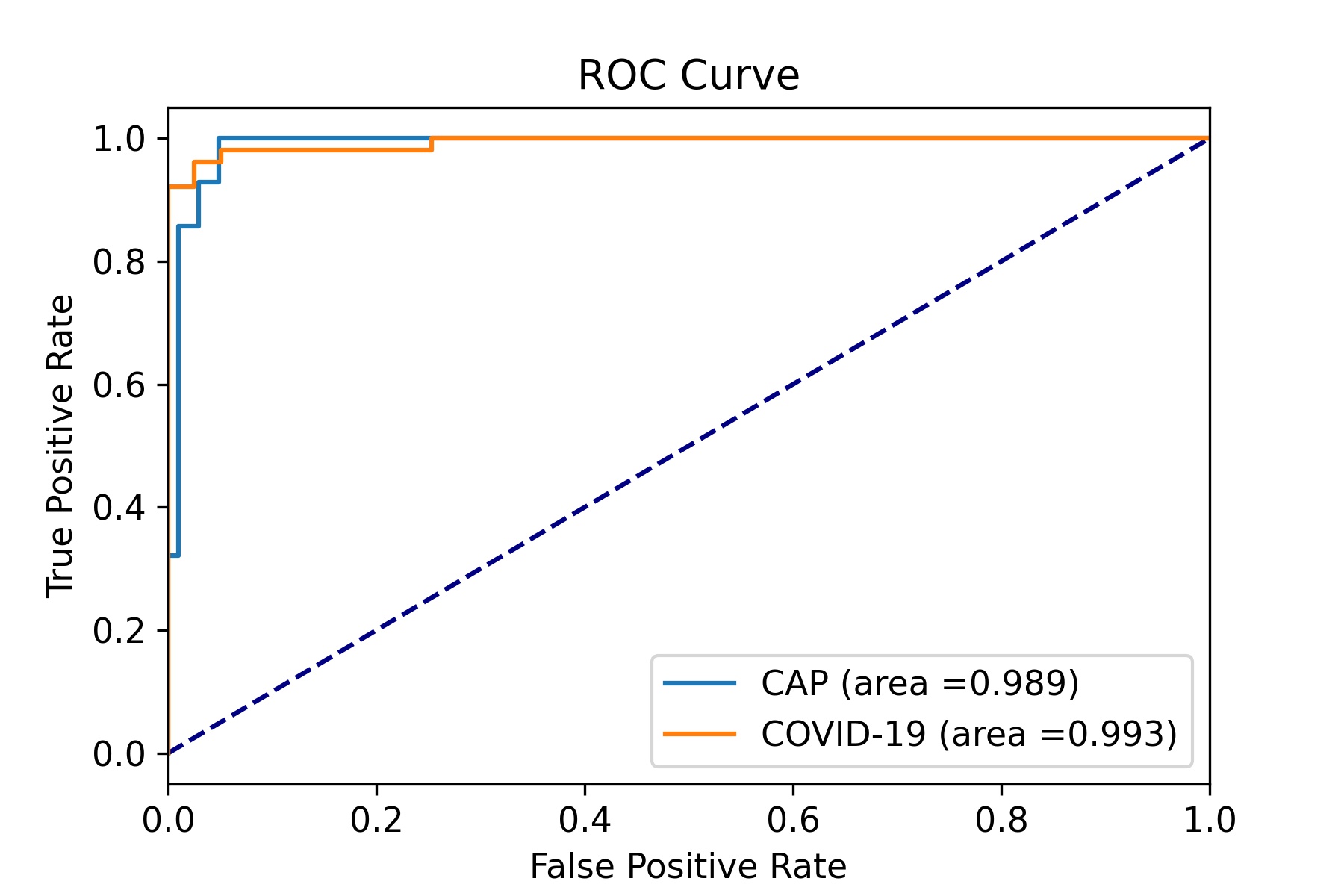}
\caption{ROC curves for COVID-19 vs. others and CAP vs. others.}
\label{fig:roc}
\end{figure}

In Fig.~\ref{fig:roc}, the ROC curves for COVID-19 and CAP cases against other classes (e.g., COVID-19 vs. CAP and Normal) are plotted. The associated AUC values are also provided.

\vspace{.1in}
\noindent
\textbf{Comparison:}
We have compared our proposed framework with the top six models~\cite{Chaudhary2021, Yang2021, Garg2021, Xue2021, Bougourzi2021, Bingyang2021} developed in the SPGC-COVID challenge.  The Methods section of this paper provides a detailed description of each model,  along with an overview of the development and evaluation steps of the challenge. In addition to the models proposed in the challenge,  we have further compared our proposed framework with another model, which utilizes the same train and test sets (excluding the 4th test set) to target the same classification task~\cite{Nicolas2021}. A brief description of this model is also provided in the Methods section.

Experimental results demonstrate that our proposed framework outperforms its counterparts proposed in the SPGC-COVID challenge. Furthermore,  it benefits from a scalable enhancement approach that can be integrated into most of the state-of-the-art models to improve their performance when testing on a heterogeneous dataset.

With regards to the development and evaluation process of the SPGC-COVID challenge,  it is worth mentioning that the proposed framework and those models whose results are used as comparison in this study have been developed in an entirely similar fashion. More specifically, the developers had access to the same datasets and labels, even the initial train/validation split of the data was the same. Moreover, no specific restrictions were applied to the development process to prevent privileging a specific type of model. In other words, the proposed framework was not designed to provide a baseline or reference standard for comparison in the challenge. In addition, we would like to highlight that even the proposed benchmark model without incorporation of the enhancement approach achieved a higher performance than other models, as shown in Table~\ref{tab:comparison}. Therefore, given that the same scenario has been in place for all the developers (including us), we believe that the results present a fair and reliable comparison. The performance of the investigated models is presented in Table~\ref{tab:comparison}.

Table~\ref{tab:comparison} illustrates the performance of seven automated models developed to tackle the same task as that of this study using the same train and test datasets. We have also compared the overall performance of our proposed framework with the aforementioned models using the statistical McNemar's test~\cite{McNemar1947} with the significance level of $0.05$. We tested the hypothesis that the models have the same proportion of errors on the entire test sets. The corresponding $p$-values are reported in Table~\ref{tab:comparison} and indicate that  the hypothesis is rejected for almost all the models except the first one as the corresponding $p$-value is slightly more than $0.05$. In other words, there is a significant difference in the proportion of errors between our proposed framework and six of the aforementioned models while such difference is not significant in the case of the model proposed in Ref.~\cite{Chaudhary2021}.

\begin{table}[t!]
\begin{adjustwidth}{-2.25in}{0in}
\centering
\begin{tabular}{|c|c|c|c|c|c|}
\hline
 Model & Accuracy(\%) & COVID-19 Sensitivity(\%) & CAP Sensitivity(\%) & Normal Sensitivity(\%) & \makecell{McNemar's test\\$p$-value}\\
\hline
\textbf{Ref.~\cite{Chaudhary2021}} & 90 & 86.27 & 89.28 & 94.11 & 0.07681\\
\hline
\textbf{Ref.~\cite{Yang2021}} & 88.46 & 86.27 & 89.28 & 90.19 & 0.03088\\
\hline
\textbf{Ref.~\cite{Garg2021}} & 87.69 & 88.23 & 78.57 & 92.15 & 0.00739\\
\hline
\textbf{Ref.~\cite{Xue2021}} & 85.38 & 84.31 & 82.14 & 88.23 & 0.00052\\
\hline
\textbf{Ref.~\cite{Bougourzi2021}} & 84.61 & 90.19 & 60.71 & 92.15 & 0.00073\\
\hline
\textbf{Ref.~\cite{Bingyang2021}} & 80.00 & 88.23 & 35.71 & 96.07 & 0.00005\\
\hline
\textbf{Ref.~\cite{Nicolas2021}} & 72.22 & 65.71 & 85.00 & 71.43 & 0.00002\\
\hline
\textbf{\makecell{Proposed\\(Benchmark)}} & 93.85 & 92.16 & 89.29 & \textbf{98.04} & 0.25\\
\hline
\textbf{\makecell{Proposed\\(Enhanced)}} & \textbf{96.15} & \textbf{96.08} & \textbf{92.86} & \textbf{98.04} & --\\
\hline
\end{tabular}
\caption{Results obtained by the models that utilized the same train and test sets. $P$-values for the McNemar's test with the significance level of $0.05$ are presented in the last column comparing the proportion of errors on the entire test sets obtained by our proposed framework and other models.}
\label{tab:comparison}
\end{adjustwidth}
\end{table}

Based on the key components of the models provided in Table~\ref{tab:models-details} and the results reported in Table~\ref{tab:comparison} and Table~\ref{tab:models-results}, we can conclude that using an advanced lung region extraction model such as the U-Net R231 COVIDweb can improve the performance. Moreover, pre-training and data augmentation are used in most of the models,  although such techniques were not utilized in our proposed CapsNets-based framework demonstrating the capability of Capsules to be trained using small datasets with limited data augmentation or pre-training.
Furthermore,  most models have not taken any specific measures to tackle the heterogeneity in the test cases. It is worth mentioning that although the model proposed in Ref.~\cite{Nicolas2021} used an online unsupervised learning approach to target this issue, it was not adequately trained and designed, in our opinion. This could possibly have led to its low performance.
In addition to the components outlined in Table~\ref{tab:models-details}, we would like to note that the best performing model in the challenge (i.e., Ref.~\cite{Chaudhary2021}) used a customized mechanism to focus on middle slices showing large visible lung areas that could possibly improve its capability to perform well on various cases. However, their approach is static and needs to be adopted depending on slice thicknesses to provide a dynamic slice selection.

\begin{table}[t!]
\begin{adjustwidth}{-2.25in}{0in}
\centering
{\color{black}
\begin{tabular}{ | m{3.6em} | m{4.8em} | m{3.2em}| m{3.2em}|m{3.2em}|m{3.2em}|m{3.2em}|m{3.2em}|m{3.2em}| }
\hline
Test Set & Sensitivity & Ref.~\cite{Chaudhary2021} & Ref.~\cite{Yang2021} & Ref.~\cite{Garg2021} & Ref.~\cite{Xue2021} & Ref.~\cite{Bougourzi2021} & Ref.~\cite{Bingyang2021} & Ref.~\cite{Nicolas2021} \\
\hline
Test1 &COVID-19 \hspace{35pt} Normal\hfill & 12/15 \hspace{35pt} 14/15 & 13/15 \hspace{35pt} 14/15 & 12/15 \hspace{35pt}15/15 & 13/15 \hspace{35pt}15/15 & 14/15 \hspace{35pt}14/15 & 12/15 \hspace{35pt}15/15 & 11/15 \hspace{35pt}12/15\\
\hline
Test2 &COVID-19 \hspace{35pt} CAP \hspace{35pt} Normal\hfill & 9/10 \hspace{35pt} 8/10 \hspace{35pt} 9/10 & 8/10\hspace{35pt} 10/10 \hspace{35pt} 7/10 & 9/10 \hspace{35pt} 8/10 \hspace{35pt} 6/10 & 9/10 \hspace{35pt} 7/10 \hspace{35pt} 7/10 & 8/10 \hspace{35pt} 7/10 \hspace{35pt} 9/10  & 10/10 \hspace{35pt} 0/10 \hspace{35pt} 10/10 & 3/10 \hspace{35pt} 9/10 \hspace{35pt} 4/10 \\
\hline
Test3 &COVID-19 \hspace{35pt} CAP \hspace{35pt} Normal\hfill & 9/10 \hspace{35pt} 10/10 \hspace{35pt} 10/10 & 9/10\hspace{35pt} 7/10 \hspace{35pt} 10/10 & 10/10 \hspace{35pt} 10/10 \hspace{35pt} 10/10 & 7/10 \hspace{35pt} 9/10 \hspace{35pt}10/10 & 10/10 \hspace{35pt}2/10 \hspace{35pt} 9/10 & 9/10 \hspace{35pt}7/10 \hspace{35pt} 9/10 & 9/10 \hspace{35pt}8/10 \hspace{35pt} 9/10\\
\hline
Test4 &COVID-19 \hspace{35pt} CAP \hspace{35pt} Normal\hfill & 14/16 \hspace{35pt} 7/8 \hspace{35pt} 15/16 & 14/16\hspace{35pt} 8/8 \hspace{35pt}15/16 & 14/16 \hspace{35pt} 4/8 \hspace{35pt} 16/16 & 14/16 \hspace{35pt} 7/8 \hspace{35pt} 13/16 & 14/16 \hspace{35pt} 8/8 \hspace{35pt} 15/16 & 14/16 \hspace{35pt} 3/8 \hspace{35pt} 15/16 & NA \hspace{35pt} NA \hspace{35pt} NA \\
\hline
\end{tabular}}
\caption{The ratio of correctly classified cases over total cases in each class obtained for the top six models that were developed based on the SPGC-COVID dataset following the Signal Processing Grand Challenge (SPGC) on COVID-19 diagnosis.}
\label{tab:models-results}
\end{adjustwidth}
\end{table}

\section*{Discussion}
In this paper, we expanded the fully-automated framework developed in our previous study~\cite{Heidarian2021} to tackle the three-way classification task (i.e., identification of COVID-19, CAP, and Normal cases) based on volumetric CT scans acquired from multiple centers using different imaging protocols. We also proposed an unsupervised enhancement approach, which can enable all deep learning-based frameworks to be adapted with the heterogeneity in different test sets. In Table~\ref{tab:confident-slices}, the numbers of slices extracted from each test set to augment the train set are presented. The low number of normal slices demonstrates the high performance of the first stage in identifying slices with and without the evidence of infection. As another advantage of the proposed framework, we can mention the capability of the Capsule Network-based model to be trained using a relatively small dataset, which is of utmost importance in the field of Medical Image Processing, in particular the COVID-19 disease, where, typically, small annotated datasets are available. The other noteworthy advantage is that the model does not require any infection annotation, which is a challenging and time-consuming task. The only segmentation used in our study is the lung area segmentation (i.e., extracting the lung parenchyma using a pre-trained U-Net model~\cite{Hofmanninger2020}), which is a well-studied task and does not add much complexity to the model.

\begin{table}[t!]
\centering
\begin{tabular}{|c|c|c|c|}
\hline
  & COVID-19 & CAP & Normal\\
\hline
\textbf{Test 1} & 595 & 0 & 4  \\
\hline
\textbf{Test 2} & 382 & 563 & 3 \\
\hline
\textbf{Test 3} & 427 & 341 & 2 \\
\hline
\end{tabular}
\caption{The number of slices extracted from each test set to augment the train set.}
\label{tab:confident-slices}
\end{table}

We would like to highlight the effect of the suggested $3\%$ threshold used to identify normal cases based on the outcome of the first stage. As mentioned earlier, $3\%$ is a safe threshold to identify normal cases as it is extremely rare to observe less than $3\%$ involvement of the lung parenchyma in COVID-19 cases. However, it is possible that the number of slices identified as infectious in a normal case exceeds this $3\%$ threshold. This could happen mainly in those CT scans with a large slice-thickness and fewer slices (e.g., less than 100 slices). In such cases, a minor error (a few number of misclassified slices by the first stage) will mistakenly indicate a large involvement of the lung parenchyma. Such errors can be avoided by increasing the $3\%$ threshold or using an adaptive threshold (e.g., based on the slice-thickness and number of slices) when we are dealing with a fewer number of slices per patient. In this study, only one normal case has been misclassified and increasing the threshold to $6\%$ could remove the error while the other cases were not affected. The promising results and benefits of the first stage in identifying slices demonstrating infection indicate its significant potential to be used in other CT scan-related models to help identify normal cases and concentrate only on a subset of slices rather than the whole volume.

Furthermore, we would like to highlight that the results shown in Table~\ref{tab:ind-result} demonstrate the incapability of the model enhanced based on a test set to improve the performance of the model on the same set. This is mainly because of the fact that the additional data used to update the benchmark model is constructed by the cases with the highest probability scores (whether correct or not) and incorporating them into the train set will force the model to further increase the corresponding probability scores while does not have much effect on other slices. As such, in the test phase, it is more reasonable to aggregate the outputs obtained by all enhanced models except the one associated with the target test set. It is also worth mentioning that due to the nature of the data (i.e., Medical Images), obtaining a large and diversified dataset from different countries is challenging. However, we will continue to expand the diversity of the dataset to perform more comprehensive investigations on the generalizability of our proposed framework on other test sets as well as determining the maximum level of the shift in image characteristics that can be compensated using our proposed framework.

Finally, it is worth noting that it is possible to design more advanced techniques to select the cases and images from the new test sets using the metrics introduced in the field of Active Learning~\cite{Smailagic2018,Smailagic2020} through which the cases which bring more diversity to the training set and the associated feature maps are detected and used for training purposes. In addition to the enhancement techniques in the field of Active Learning, there have been recently several studies on using Generative Adversarial Networks (GANs) to cope with the data and domain shift in medical images~\cite{Tzeng2017,Pandey2020} where the labeled data is not available in the target domain. The main goal in such frameworks is to achieve a domain invariant image representation which can efficiently embed the important features of the image regardless of the imaging modality or imaging technique. Similarly in~\cite{Ahn2020}, an auto-encoder and feature augmentation-based approach is proposed to adapt the model with various imaging modalities obtained by different scanners. However, in this study, we are dealing with only one imaging modality (i.e., CT scan) and the level of characteristic shift between the images is lower compared to the images investigated in the aforementioned studies. Moreover, we could achieve high performances using a far less complicated mechanism.

In summary, we have proposed an approach to update the model's parameters by extracting confident predictions from the test sets and utilize them to re-train the model in order to increase its capability and robustness in the presence of gaps between the imaging protocols and patients' clinical history. We showed that we can train different versions of the model based on different test sets and combine their outputs to generate the final predictions, which are more accurate and robust.

As a final note to our discussion, a technical review on the current clinical significance of the chest imaging, in particular CT scans, in COVID-19 diagnosis during pandemic is provided. Furthermore, we provide an overview of new techniques introduced to minimize the associated cumulative radiation dose imposed on the body to address the concerns raised around the potential risks of CT imaging.
\vspace{0.1in}
\\
\noindent
\textbf{Clinical Significance of Chest Imaging and CT in Screening of COVID-19 }
\vspace{0.1in}
\\
\noindent
First of all, we know that the role of chest imaging, especially CT, has evolved during the pandemic following the accumulation of experience and scientific data. The performance of chest imaging has been debated since the early period of the pandemic, where early studies from China showed a superiority of CT over the RT-PCR test~\cite{Inui2021,Machnicki2021}. This might have been attributed to variabilities in viral load depending on the disease stage and sampling error, as well as the low availability and high demand of the test in the early days of the pandemic~\cite{Malguria2021}.
In this regard, some studies have recommended parallel testing using CT and RT-PCT, especially when consecutive tests are required to confirm the infection for treatment planning~\cite{Wang2020,AlTawfiq2020,Sun2020}.
Currently, RT-PCR test is the most commonly used diagnostic tool for COVID-19 detection, and some scientific societies do not recommend the use of chest CT for screening of COVID-19. On the other hand, chest imaging, specifically CT scanning, has a crucial role in different healthcare environments and clinical scenarios~\cite{Rubin2020}.
Although it is not recommended for asymptomatic or mildly symptomatic patients in the absence of accompanying risk factors in an environment, which is abundant in resources, unless they are at risk for disease progression, chest CT is recommended for medical triage of patients with suspected COVID-19 who present with moderate to severe symptoms and a high pretest probability of the disease regardless of the RT-PCR test, or in resource-constrained environments where RT-PCR tests may not be readily available or test results might be delayed. Chest CT imaging is also recommended for patients with worsening respiratory status. As such, it is clear that despite the widespread use of the RT-PCR test for screening of COVID-19 infection, the role of chest CT is irreplaceable in providing a baseline for future comparison, revealing an alternative diagnosis, establishing manifestations of important comorbidities in patients with risk factors for disease progression, and influencing treatment strategy and the intensity of monitoring for clinical deterioration~\cite{Inui2021}.

In addition, it should be noted that the RT-PCR test is accompanied by a high false negative rate. Based on the latest guidance provided by the World Health Organization (WHO) for critical preparedness, readiness, and response actions for COVID-19~\cite{WHO2021}, such errors most likely occur due to technical reasons such as new virus mutations, the inhibition of the PCR reaction, sampling or storage errors, and timing of the test (i.e., too early or too late examinations).
It is also worth noting that although a consecutive RT-PCR test could reduce the error probability, several studies~\cite{Hao2020,Bouiller2020} have shown that such false negative results are not limited to the first test, and multiple negative PCR results have been reported for many cases while significant progression of the disease has been confirmed through their CT scans.

In summary, the rt-PCR test is currently considered as the most reliable tool for screening for COVID-19.  As stated above, there are certain clinical scenarios,  which can include a significant number of cases, where the role of CT in diagnosing COVID-19 infection is irreplaceable. These scenarios include : (i) Suspected false negative PCR test, i.e., moderately or severely symptomatic patients with a high pretest probability of the disease regardless of the rt-PCR test; (ii) Inavailability of PCR test, i.e., resource-constrained environments,  where the rt-PCR test may not be available, and; (iii) Latency, i.e., the rapidity with which the information is provided with chest CT compared to an rt-PCR test makes it the preferred method of diagnosis in certain environments. In conclusion, although we believe that there is a critical role for chest CT in correctly identifying COVID-19 infection in addition to being a critical approach for detection of complications and prognosis, there are concerns on limited clinical significance of chest CT as a tool for diagnosis of COVID pneumonia.

\vspace{0.1in}
\textbf{Radiation Dose Reduction in CT Imaging}
\vspace{0.1in}
\\
\noindent
As the next part of our discussion, we would like to address the concerns raised around the potential risks of radiation exposure caused by CT imaging and its side-effects on the patients' bodies. In particular, we would like to highlight the recent recommendations in utilization of low-dose and ultra-low-dose CT imaging protocols, which ensure that minimized radiation is imposed on the body while the images are still of high quality and could reveal specific radiologic findings. Several studies have introduced specific technical settings that result in a significant reduction in the associated CT dose
index (CTDIvol). As an example of such studies, Reference~\cite{Kang2020} reported adequate assessment of pulmonary opacities related to COVID-19 pneumonia at $100$ kV
with tin filter and iterative reconstruction technique with a CTDIvol of $0.4$ mGy versus the standard-dose protocol with the CTDIvol of $3.4$ mGy. Another study~\cite{Kang2020} applied $100$ kV with tin filter
and $0.6$ second exposure time using a high pitch and fast gantry rotation time to acquire chest CT images at $0.6$ mGy CTDIvol, which were comparable to standard-dose chest CT at $6.4$ mGy.
Furthermore, we would like to highlight that the low-dose scanning protocol used to acquire the CT scans in test set 1 and 4 of this study has resulted in a significant reduction in the average radiation dose by reducing the reference mAs value from $50$ to $15-20$ . More specifically, the radiation dose was reduced from the estimated value of $7$mSv in standard-dose scans to $1-1.5$mSv in LDCT scans, and $0.3$mSv in the ULDCT ones, which is as low as that of a single chest radiograph.
The results obtained for the aforementioned test sets demonstrate the effectiveness of such scans in providing specific CT findings that are efficiently captured by the proposed framework in an automated fashion.
In conclusion, the concerns around the cumulative radiation exposure on patients' bodies could be mitigated to some extent, especially during a pandemic when a larger population is in need of being scanned.

\section*{Acknowledgments}
This work was partially supported by the Natural Sciences and Engineering Research Council (NSERC) of Canada through the NSERC Discovery Grant RGPIN-2016-04988 and NSERC Discovery Grant RGPIN 2019 06966. We would also like to express our appreciation to all the teams participated in the SPGC-COVID challenge,  especially the teams whose approaches are summarized and cited in this paper for their efforts in proposing novel ideas and enhancing the challenge's quality. 

%
%
%

\end{document}